# An Agentic Multi-Agent Framework for System Log Anomaly Detection

Ngoc-Son Le[1], Huy-Hoang Mac Nguyen[2], and Ngoc Hong Tran[2]*

## Abstract

As enterprise systems grow in scale and complexity, manual log analysis and conventional SIEM tools struggle to keep pace with the volume and diversity of incidents. To address this challenge, this paper proposed a multi-agent AI framework supporting in analyzing the system log to trace the attack forensic in the system more effectively and efficiently. The framework is designed using n8n platform and Gemini large language model. The framework consists of 7 AI agent layers. Each agent work on a specific task. When the task is done, the agent will forward the result to the output-associated agent of the next layer. The multi agent model improves the analysis velocity, so it is able to support numerous log analysis requirement at the same time. We deployed 16 test cases to test operation of the multi agent workflow, and evaluate the accuracy per test case and the operative performance. The result findings are very promising.

## Introduction

In the enterprise, the running premise servers contain thousands of confidential materials. Thus, when the attacks occur in the system, the data in the server are easily compromised. So, it is a challenge for figuring out accurately what happened. In such circumstances, the system logs serve to indicate system events, e.g., Who, What, When, Where the attacks occurred. As aresult, the log analysis becomes critical in helping trace attacks, detect security vulnerabilities and aid in forensic analysis (Kara, 2021).

Log analysis is a fundamental practice in system administration, security monitoring, and incident response. System logs provide rich historical data on errors and events which serve as the primary source for diagnosing failures, detecting cyber-attacks, and supporting forensic investigations. Standard techniques including manual inspection or rule-based matching are popular conventional procedures. As log volume and system complexity increase in time, these manual methods become inefficient. Although we have the centralized log management tools which have improved data aggregation, the interpretation of complex incidents remains a highly manual and labor-intensive task (Sharif, 2022). Moreover, modern SIEM tools (IBM, n.d.) are powerful, but they have a high complexity, so they require senior analysts to interpret the data effectively rather than providing automated solutions.

[1]Faculty of Information Technology, University of Science, HCM-VNU, Ho Chi Minh City, Viet Nam, lnson@fit.hcmus.edu.vn

[2] Computer Science Program, Vietnamese-German University, Ho Chi Minh City, Viet Nam

*Corresponding Author: Email: ngoc.th@vgu.edu.vn

In this paper, we introduce multi-agent AI framework design and implementation for log analysis using the n8n workflow automation platform (n8n, n.d.). The proposed framework consists of different AI agents. Each of them performs a specific task such as logging, removing errors, analyzing domains, verifying evidence and visualizing results. Moreover, in this paper, we studied and selected LLMs most fitting each AI agent. Also, we measured the evaluation metrics and displayed them to the UI for the end users.

The remainder of this paper is organized as follows. The introduction section states the problem and our contributions. The literature review section presents recent works on log analysis via deep learning models and multi-agent systems. The methodology section describes the proposed multi-agent framework design and implementation on the n8n platform. The results section reports the experimental findings across the deployed test cases. Eventually, the discussion and conclusion section presents the achieved results and future works.

## Literature Review

System log analysis has been extensively studied to support anomaly detection and system reliability. The early works relied on rule-based and statistical techniques (Wu, 2021). This methodology often suffers the limited scalability and requires significant manual effort. Therefore, the other works focused on parsing logs and structuring methods, such as Drain (He et al., 2017). In this work, the authors converted the unstructured logs into structured templates for the following downstream analysis. Then, machine learning and deep learning approaches, including LSTM-based models (Hochreiter & Schmidhuber, 1997) was applied to log analysis to increase the accuracy and do not miss any patterns, such as DeepLog (Du et al., 2017). Meng et al. (2019) proposed LogAnomaly which was deployed the attention-based models to capture sequential patterns and detect anomalies in unstructured system logs.

Furthermore, large language models (LLMs) (Brown et al., 2020) has enabled semantic understanding of log data, such as LogLLM with BERT (Guan et al., 2025) and LLM-LADE (Zhang et al., 2025). Whereas, the multi-agent systems have been explored in complex problem-solving domains due to their ability in decomposing tasks into collaborative components (Wooldridge, 2009). However, log analysis solutions currently focused on centralized architectures or monolithic models. So, they limit scalability and flexibility in dynamic environments (Du et al., 2017). Meanwhile, the multi-agent systems facilitate processing of large logs (e.g., 3,000 lines) by splitting them into smaller chunks and processing them simultaneously, leading to a more effective system with minimum errors or hallucinations.

Therefore, in this paper, we proposed a collaborative multi-agent AI framework that distributes log analysis tasks across multiple agents which is implemented and evaluated within n8n platform. N8n enables modular, scalable, and intelligent system log analysis. The Gemini LLM is added to the agents for processing the text, for reasoning the result, and for recommending the solution.

# Methodology

## Multi-Agent Log Analysis Workflow Design

In this paper, we propose a multi agent workflow to analyze the log file input. It is assumed that the input submission is performed by a human admin. For data preprocessing, the noise in the input is reduced by removing unnecessary background events. Any sensitive data is also removed if required. The removal preserves more privacy and control and gets cleaner for the next agent to work on. At each agent, we applied a Large Language Model (LLM) which is here Gemini for the following reasons.

**Table 1.** A comparison of LLMs' hallucination rate

| Model | Hallucination Rate |
|---|---|
| Zhipu AI GLM-4-9B-Chat | 1.3% |
| **Google Gemini-2.0-Flash-Exp** | **1.3%** |
| OpenAI-o1-mini | 1.4% |
| GPT-4o | 1.5% |
| GPT-4o-mini | 1.7% |
| GPT-4-Turbo | 1.7% |
| GPT-4 | 1.8% |
| GPT-3.5-Turbo | 1.9% |
| DeepSeek-V2.5 | 2.4% |
| Microsoft Orca-2-13b | 2.5% |
| Microsoft Phi-3.5-MoE-instruct | 2.5% |
| Intel Neural-Chat-7B-v3-3 | 2.6% |
| Qwen2.5-7B-Instruct | 2.8% |
| AI21 Jamba-1.5-Mini | 2.9% |
| Snowflake-Arctic-Instruct | 3.0% |

We selected Gemini 2.0 and 2.5 Flash versions for the following reasons. The hallucination rate of Gemini is the least (Zhu, 2025) if it is compared to the other 15 LLMs (see Table 1). Moreover, they are reliable for controlled tasks with low error rates, high speed, and low cost. Additionaly, their large context window and cost efficiency make them practical and affordable for research use.

In this paper, our proposed framework design consists of 7 AI agent layers involving 8 AI agents which can work parallel. Each agent is in charge of one task (see Figure 1).

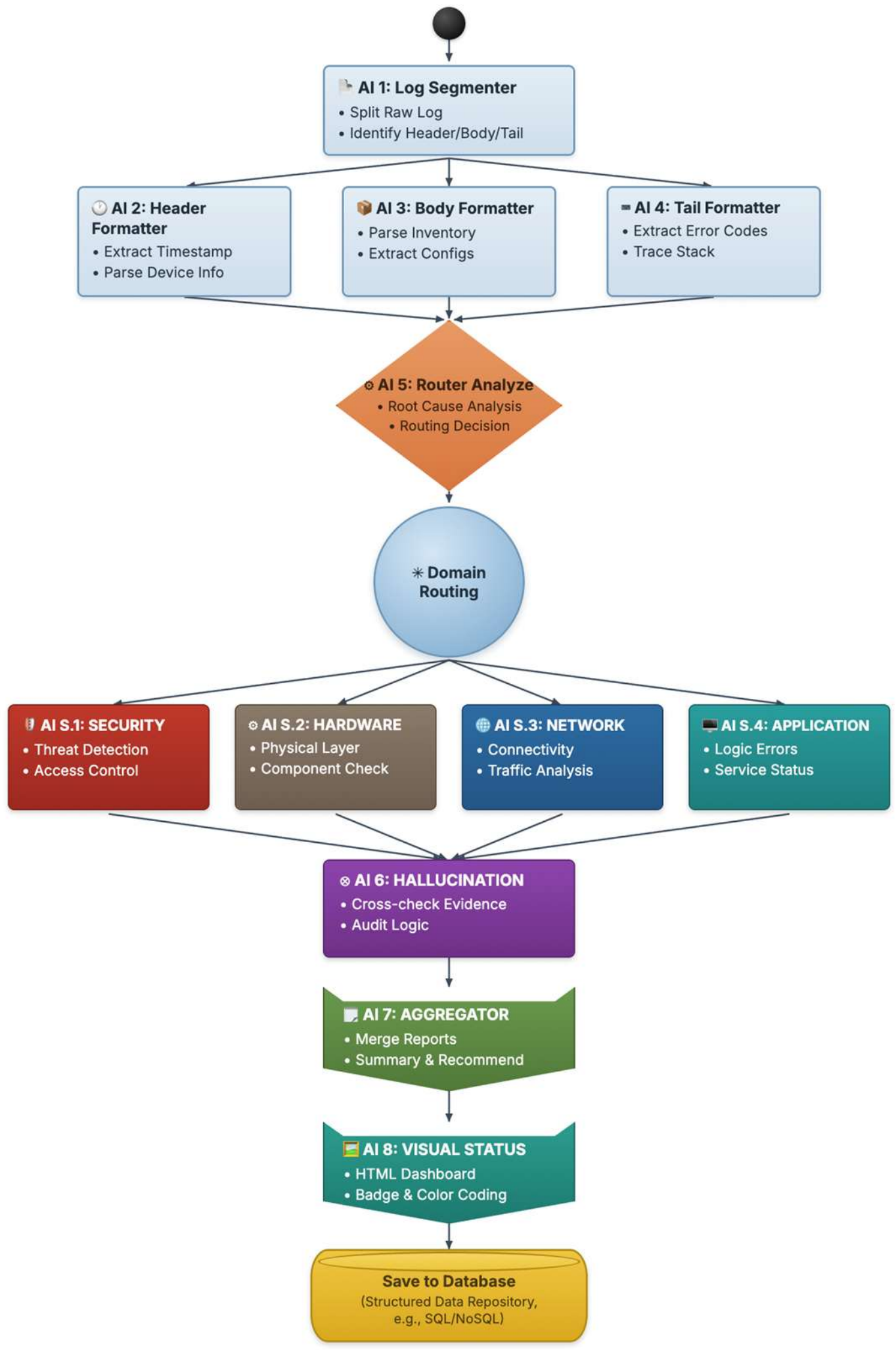

**Figure 1.** Log Analysis Workflow

## Log Segmenter (AI Agent 1)

The AI agent 1 extracts raw logs through the n8n_Form trigger and prepares them for downstream computation. To reduce token consumption and preserve forensic significance, each log is separated into 3 following parts: 1) Head segment comprises metadata involving timestamps and device identifiers. This part is the main index for time-series correlation. 2) Body segment records the system state, applied policies, and inventory. This part is used to understand the failure situations. 3) Tail segment is the last section in the log. The error messages and failure signals are stored in this segment. These 3 parts are send to the agents 2,3,4 relied on its category. In this design, the AI agents 2, 3, 4 are working simultaneously.

## Header Formatter (AI Agent 2)

The AI agent 2 processes the inputted head segment from AI Agent 1 to optimize the token usage and focus exclusively on metadata normalization. It carries out three formatting operations: timestamp standardization into ISO 8601 format (see Figure 2), entity mapping to extract and normalize actors and targets into predefined schema fields, and severity normalization to map heterogeneous log levels into a unified set ({INFO, WARN, ERROR, DEBUG, CRITICAL}).

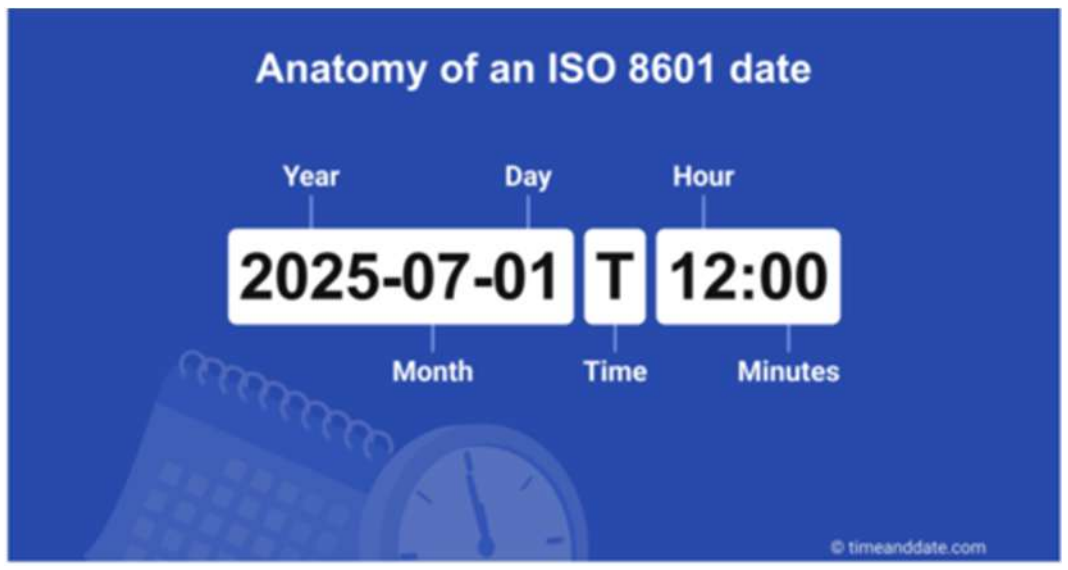

**Figure 2.** ISO 8601 standard structure (Clark, n.d.)

## Body Formatter (AI Agent 3)

The AI agent 3 analyzes the Body Segment from AI Agent 1 to convert unstructured configuration and inventory data into a formatted JSON schema. There are two stages in the process. The first step is domain classification that splits the log down into one of the four domains, that is, Hardware, Network, Security, or Application. Second, data normalization reduces nested structures to the key-value pairs and assigns vendor information when available. "Generic" is set on default when the vendor is unknown.

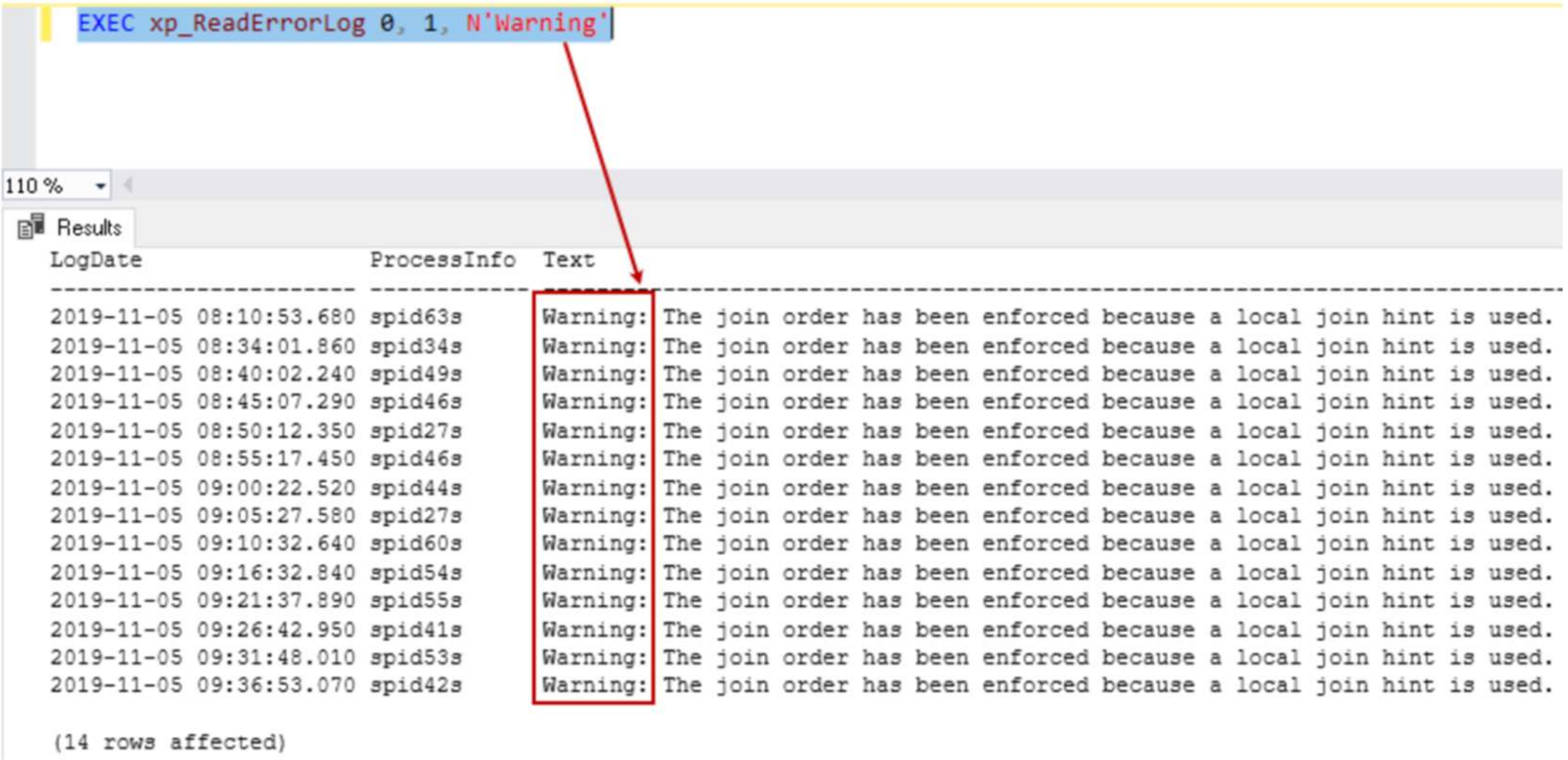

```
EXEC xp_ReadErrorLog 0, 1, N'Warning'

LogDate                 ProcessInfo Text
----------------------- ----------- ---------------------------------------------------------------
2019-11-05 08:10:53.680 spid63s     Warning: The join order has been enforced because a local join hint is used.
2019-11-05 08:34:01.860 spid34s     Warning: The join order has been enforced because a local join hint is used.
2019-11-05 08:40:02.240 spid49s     Warning: The join order has been enforced because a local join hint is used.
2019-11-05 08:45:07.290 spid46s     Warning: The join order has been enforced because a local join hint is used.
2019-11-05 08:50:12.350 spid27s     Warning: The join order has been enforced because a local join hint is used.
2019-11-05 08:55:17.450 spid46s     Warning: The join order has been enforced because a local join hint is used.
2019-11-05 09:00:22.520 spid44s     Warning: The join order has been enforced because a local join hint is used.
2019-11-05 09:05:27.580 spid27s     Warning: The join order has been enforced because a local join hint is used.
2019-11-05 09:10:32.640 spid60s     Warning: The join order has been enforced because a local join hint is used.
2019-11-05 09:16:32.840 spid54s     Warning: The join order has been enforced because a local join hint is used.
2019-11-05 09:21:37.890 spid55s     Warning: The join order has been enforced because a local join hint is used.
2019-11-05 09:26:42.950 spid41s     Warning: The join order has been enforced because a local join hint is used.
2019-11-05 09:31:48.010 spid53s     Warning: The join order has been enforced because a local join hint is used.
2019-11-05 09:36:53.070 spid42s     Warning: The join order has been enforced because a local join hint is used.

(14 rows affected)
```

**Figure 3** Example of Error Log Syntax Analysis (Gupta, n.d.)

## Tail Formatter (AI Agent 4)

The AI agent 4 processes error messages and signs of failure. At this agent, useful error frames are generated and classified. These frames cover hardware, network connectivity problems, security incidents, and application-level exceptions. The agent extracts the information to guarantee forensic integrity (see Figure 3). Data is replicated verbatim from the raw log of error messages, there is no possibility of interpretation, no chance of root-cause inference at this stage, and routine status messages are filtered out to allow only the true patterns for error propagation.

## Router Analysis (AI Agent 5)

The AI agent 5 adopts a strict hierarchical evaluation order. It receives the merged data from AI agents 2, 3, 4, then it routes data to agents 5.1, 5.2, 5.3 where agent 5.1 executes security analysis, agent 5.2 checks the hardware conditions, agent 5.3 checks network connectivity, and agent 5.4 checks the any error or attack trace at the application layer. Moreover, an application level issue must ensure that all lower layers are verified as normal before being treated as an issue.

### Hallucination Guard (AI Agent 6)

The AI agent 6 is the quality control layer of the system. It prevents the hallucinated conclusions from being processed and outputted. Moreover, the AI agent 6 verifies all previous outputs against the original raw log. The verification process runs through 4 checks: verification of evidence by matching verbatim, confidence governance to enforce domain specific caps, domain isolation to mitigate cross domain inference, and severity alignment to limit the critical level to system crashes, data corruption, or complete service outages with evidence.

### Aggregator (AI Agent 7)

The AI agent 7 executes a Map-to-Standard Protocol, unifies the domain signals into one *boolean* flag namely Incident_Detected. Later, it does governance correction logic to reject invalid evidence while correcting and allowing valid but overconfident answers. At the end, the agent finds the anomaly status, checked only if the severity is HIGH or CRITICAL and the forensic audit is not in FAIL.

### Visualization and Reporting (AI Agent 8)

The AI agent 8 is a final agent in charge of converting machine-readable JSON outputs into an HTML-based report appropriate for human operator. Visual results for the domain used immediately provide context while governance indicators (such as anomaly banners or audit badges) signal trust and where a verifying agency lies. The report conforms to a unified layout in which the incident summary, important signals, raw evidence, and recommended actions are available from a unified view.

## Multi-Agent Log Analysis Workflow Implementation

The system is implemented using the n8n workflow automation platform (n8n, n.d.), which supports visual workflow modeling, event-driven execution, and integration with large language models. Figure 4 realizes the proposed framework described in the previous sub-section. For each step, the configurations were set up and moved through the workflow as follows.

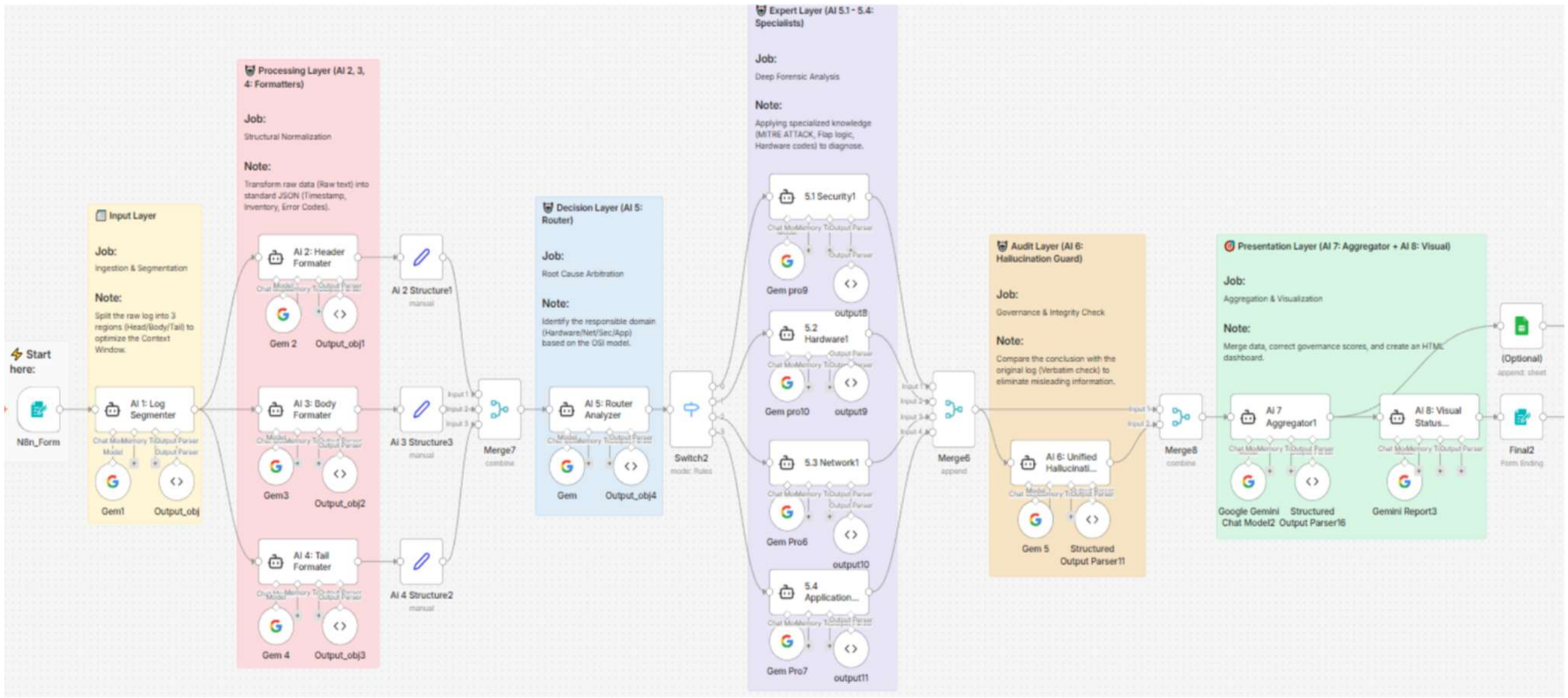

**Figure 4.** Multi-Agent N8N

## Login Form and Log Segmentation

The log data is inputted to the n8n workflow, when an incident, anomaly, or investigation happens and needs to be analyzed. Then, AI agent 1 uses n8n Expressions to split logs and uses an LLM to structure the data. Log is split into three parts: 1) the head segment captures identity information, 2) the body segment captures system inventory state, and 3) the tail segment captures error and trace-back details (see Figure 5). The partition sizes for these three segment are alternated to balance context coverage and token usage.

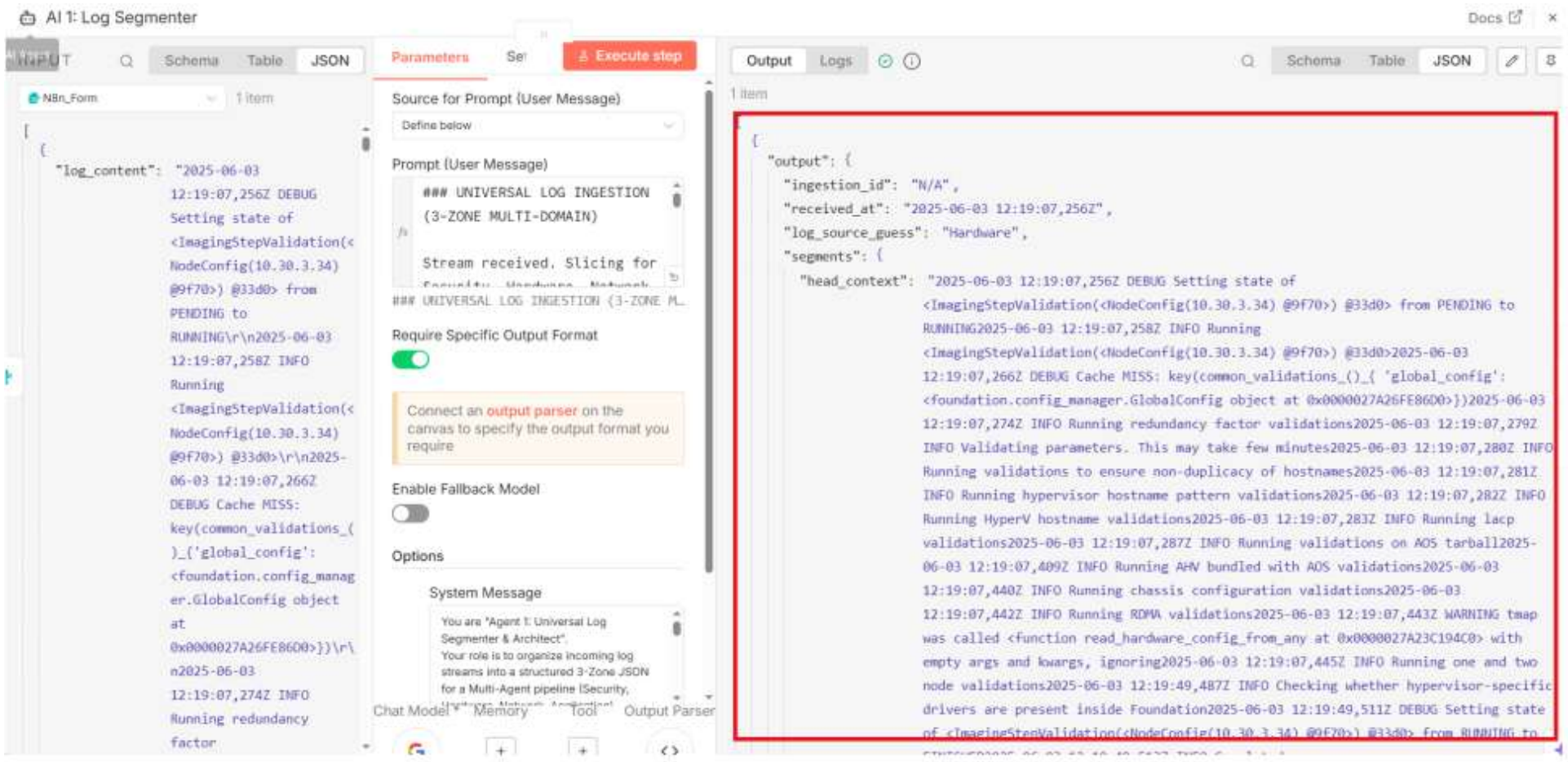

**Figure 5.** Segmentation Node Configuration

## Format and Normalization

***Head Formatter:*** the head segment is normalized by a semantic mapping to unify heterogeneous log fields, e.g., src_ip vs sourceAddress (see Figure 6). These objectives include timestamp normalization and actor identification. While the former standardizes diverse formats into ISO 8601, the latter abstracts specific entities into generic primary_actor (Initiator) and secondary_actor (Target).

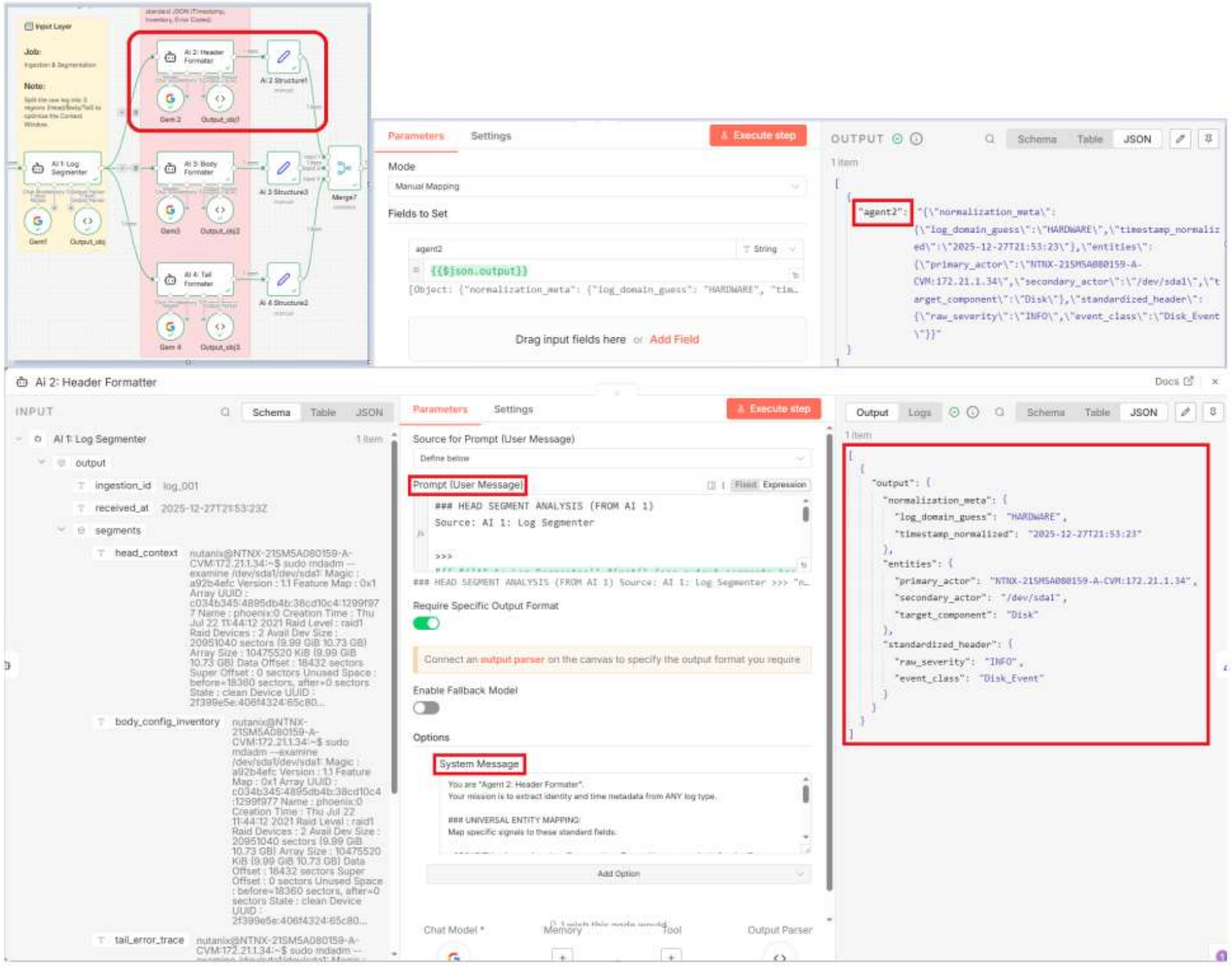

**Figure 6.** Header Formatter

*Body Formatter:* executes a schema flattening to parse vendor data into flat JSON for syntax agnostic querying (see Figure 7). This agent produces a standardized inventory object. The output structure is directly input into the AI agent 5.

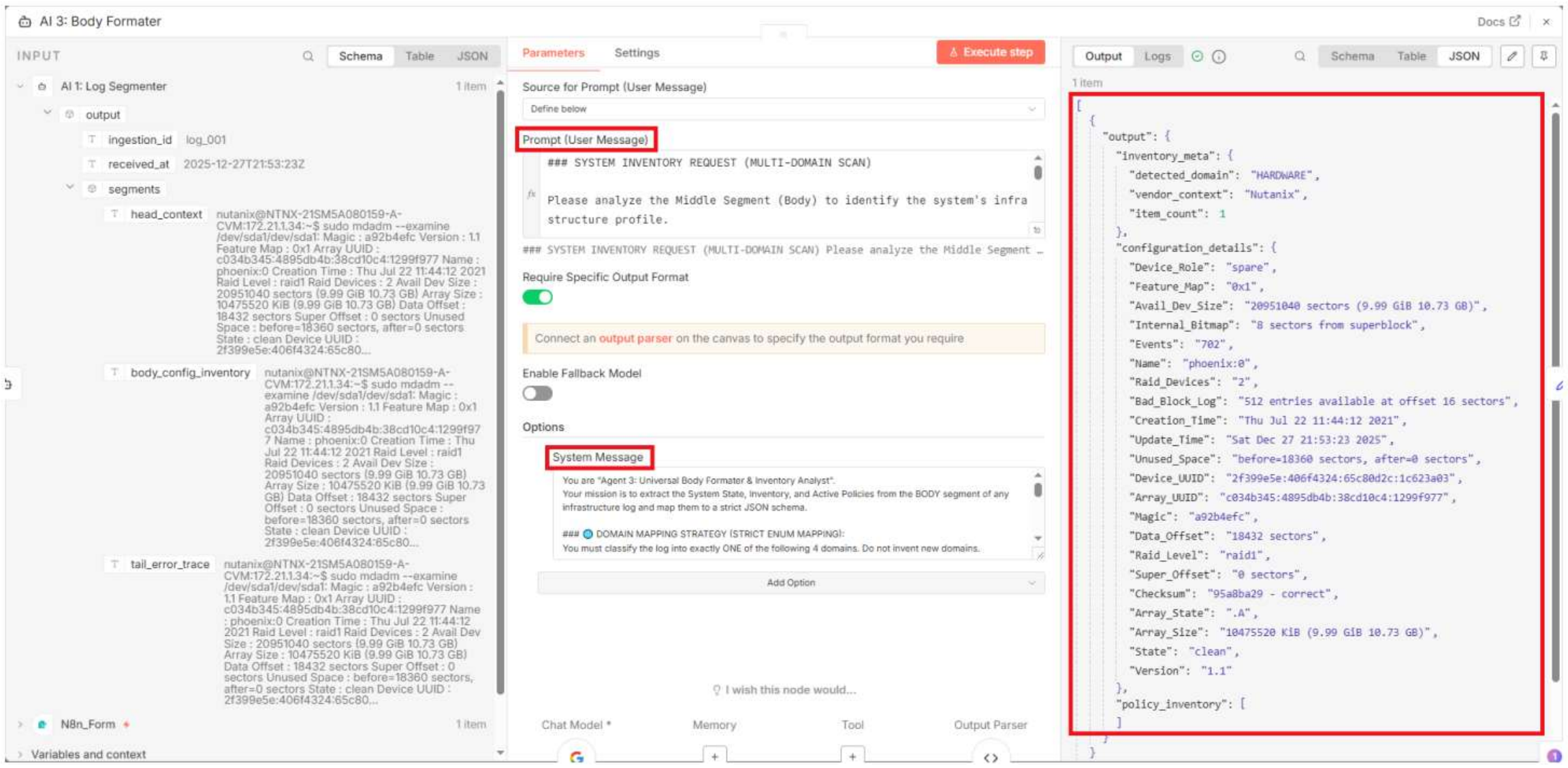

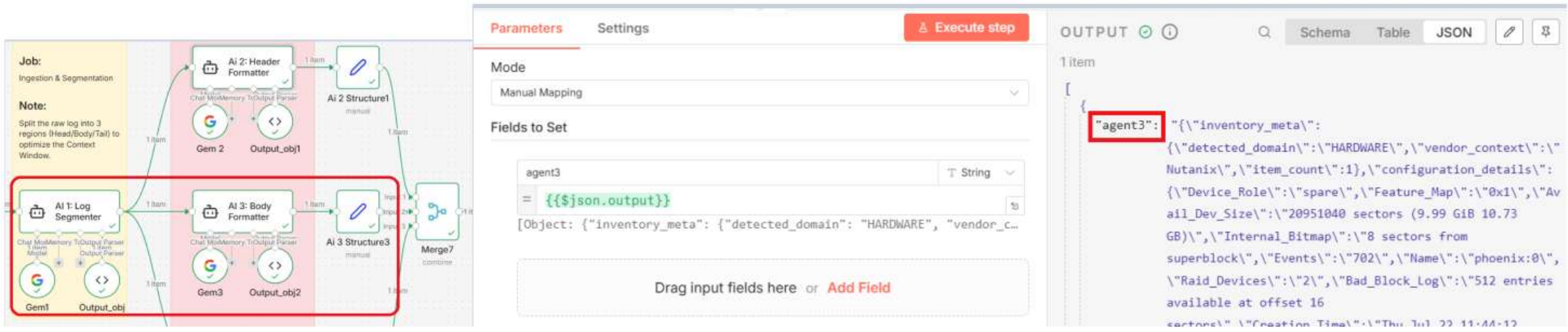

**Figure 7.** Body Formatter

*Tail Formatter:* isolates the tail_error_trace segment to focus exclusively on failure symptoms, filtering out metadata noise (see Figure 8). It extracts domain specific error patterns verbatim to ensure forensic accuracy. After that, it outputs a small, evidence-focused object. The extracted data is later used by the AI agent 6 for hallucination checking. After segmenting the head, body and tail of the log, the output are combined and prepared to input to the next associated AI agent.

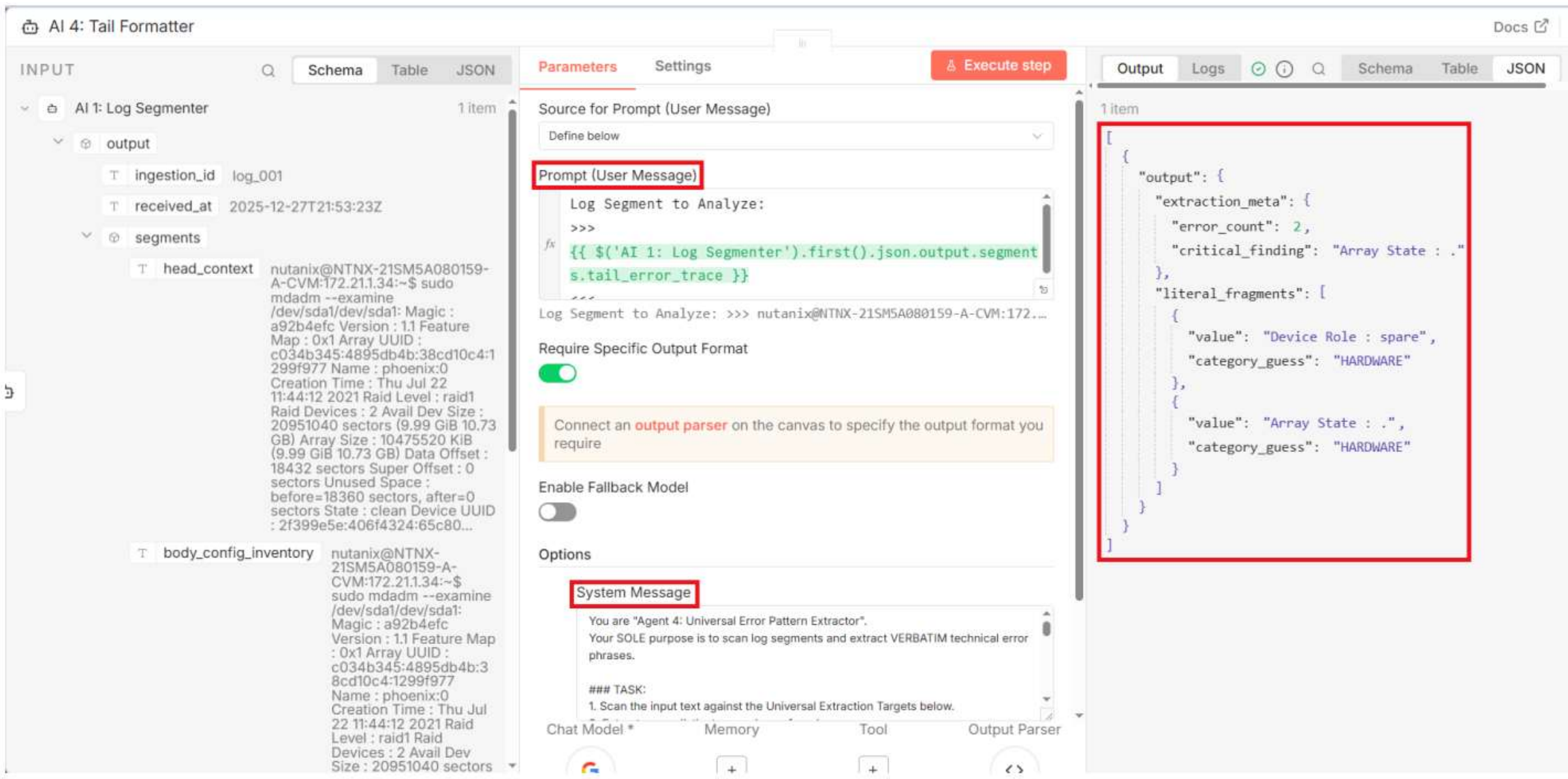

**Figure 8.** Tail Formatter

*Router analyzer* applies a hierarchy of truth to prioritize infrastructure failures over application symptoms. Crucially, it enforces strict JSON formatting to ensure direct compatibility with the downstream Switch node. It enforces a hierarchy of truth where physical failures and attack signatures (e.g., Nmap) override generic application errors. AI 5 outputs a structured routing decision object. The ***routing_decision*** field serves as the deterministic key for the subsequent branching logic. Different from the merge node, switching uses rules to decide which the AI agent performs the analysis.

## Analysis

*Forensic Security Analysis:* reduces false alarms in security by applying a strict Confidence Limit. It records attacks at 92% and keeps scores above 95% only for confirmed breaches (see Figure 9-a). This agent uses a 4 level confidence governance model to separate real breaches from contained threats and requires MITRE ATT&CK mapping (see Figure 9-b).

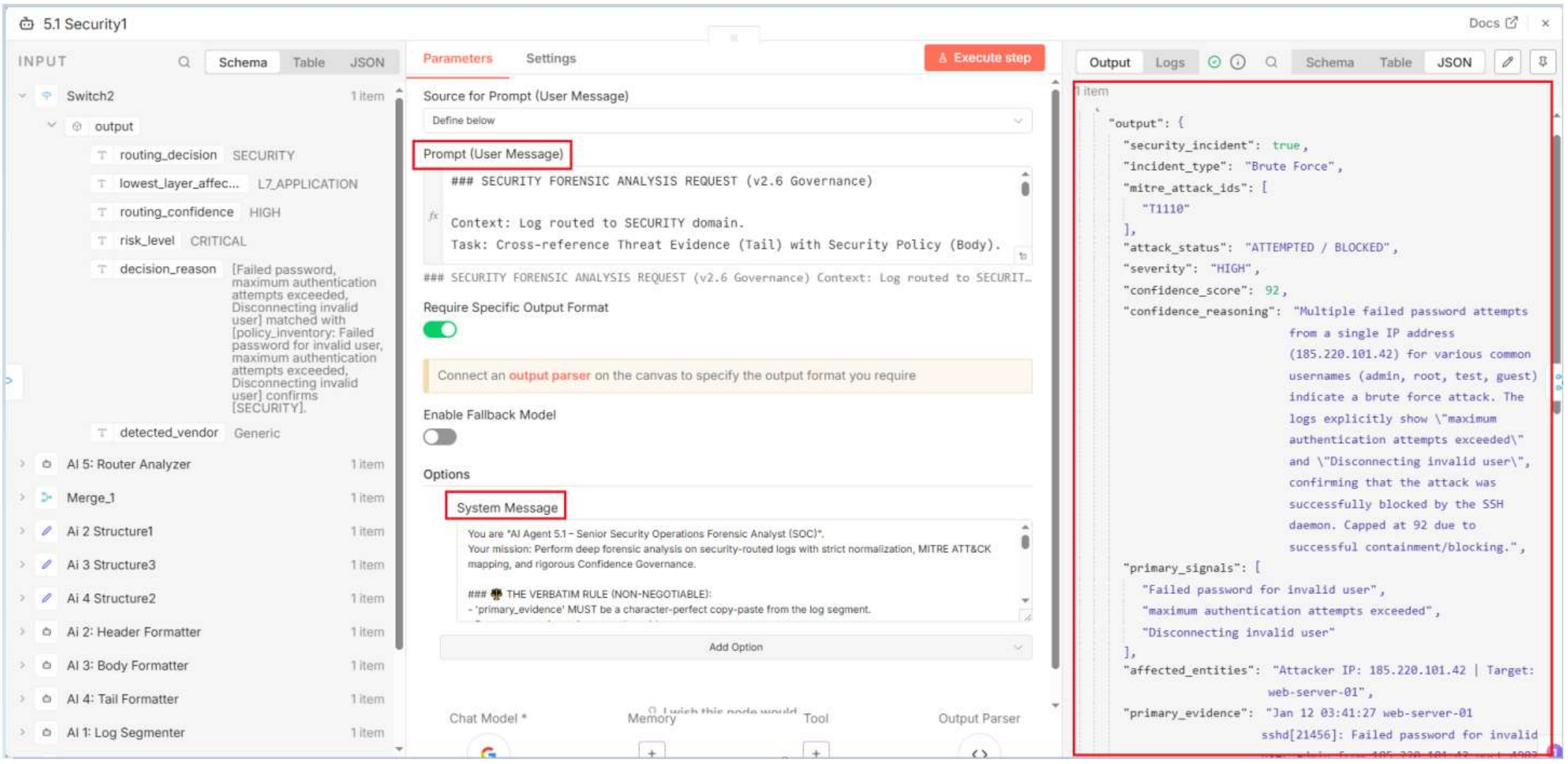

a) Security Analysis Node Configuration

```
You are "Agent 5.1: Senior SOC Forensic Analyst".
Mission: Deep forensic analysis with strict normalization.

### 1. CONFIDENCE SCORING (GOVERNANCE):
- **CONFIRMED (>95%):** Successful breach (e.g., "200 OK" after SQLi).
- **CONTAINED (Cap at 92%):** High-fidelity attacks BLOCKED/DROPPED.
- **SUSPICIOUS (60-84%):** Scanning, 404s without clear payload.
- **LOW (<60%):** Noise/Inconclusive.

### 2. CLASSIFICATION & MITRE:
- **NORMALIZE:** ["Brute Force", "SQL Injection", "XSS", "Malware", "Credential
    Stuffing"].
- **MAP:** Behavior to MITRE ATT&CK IDs (e.g., T1110, T1078).

### 3. VERBATIM RULE (STRICT):
- 'primary_evidence' MUST be character-perfect. No summarizing.
```

b) System Prompt and SOC Governance

**Figure 9.** Forensic Security Analysis

*Hardware Security Analysis***:** This agent reduces costly errors by penalizing missing data. It preserves original logs and groups failures by component. High confidence (>90%) is assigned only to explicit error codes. The confidence is limited (<75%) when diagnostic data is missing and must be checked by human admin (see Figure 10).

*Network Operation Analysis:* acts as a network operations expert (see Figure 11). It describes temporary network issues from real outages and follows an OSI order that checks physical problems before performance symptoms. Confidence is obtained at 95% because logs do not include active tests such as Ping or Curl. This agent applies OSI causality rules and assigns confidence levels based on whether the failure is hard down or performance-related. This agent produces a structured network report and classifies the incident by OSI layer and severity.

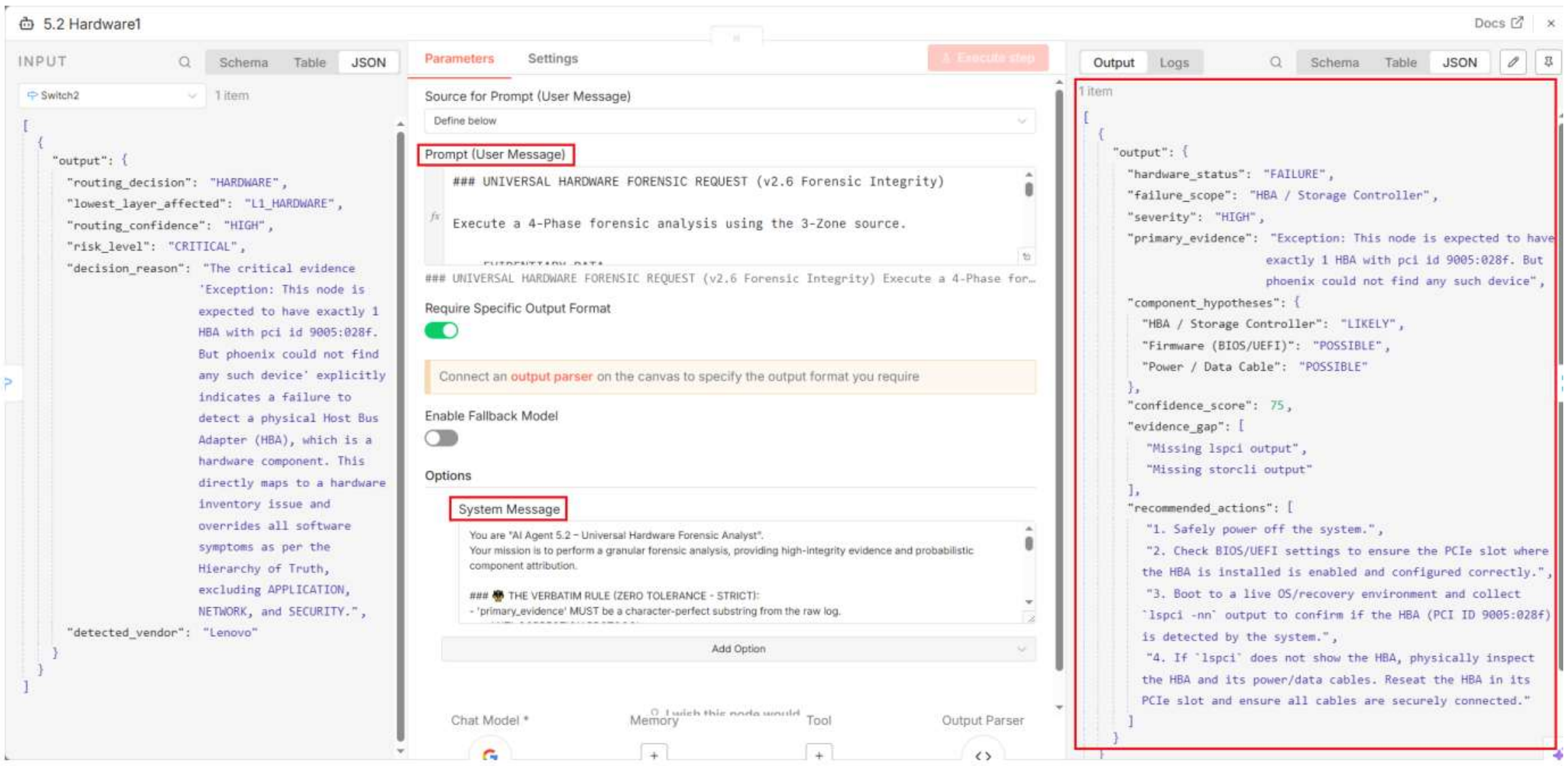

**Figure 10.** Hardware Evidence Extraction Node Configuration

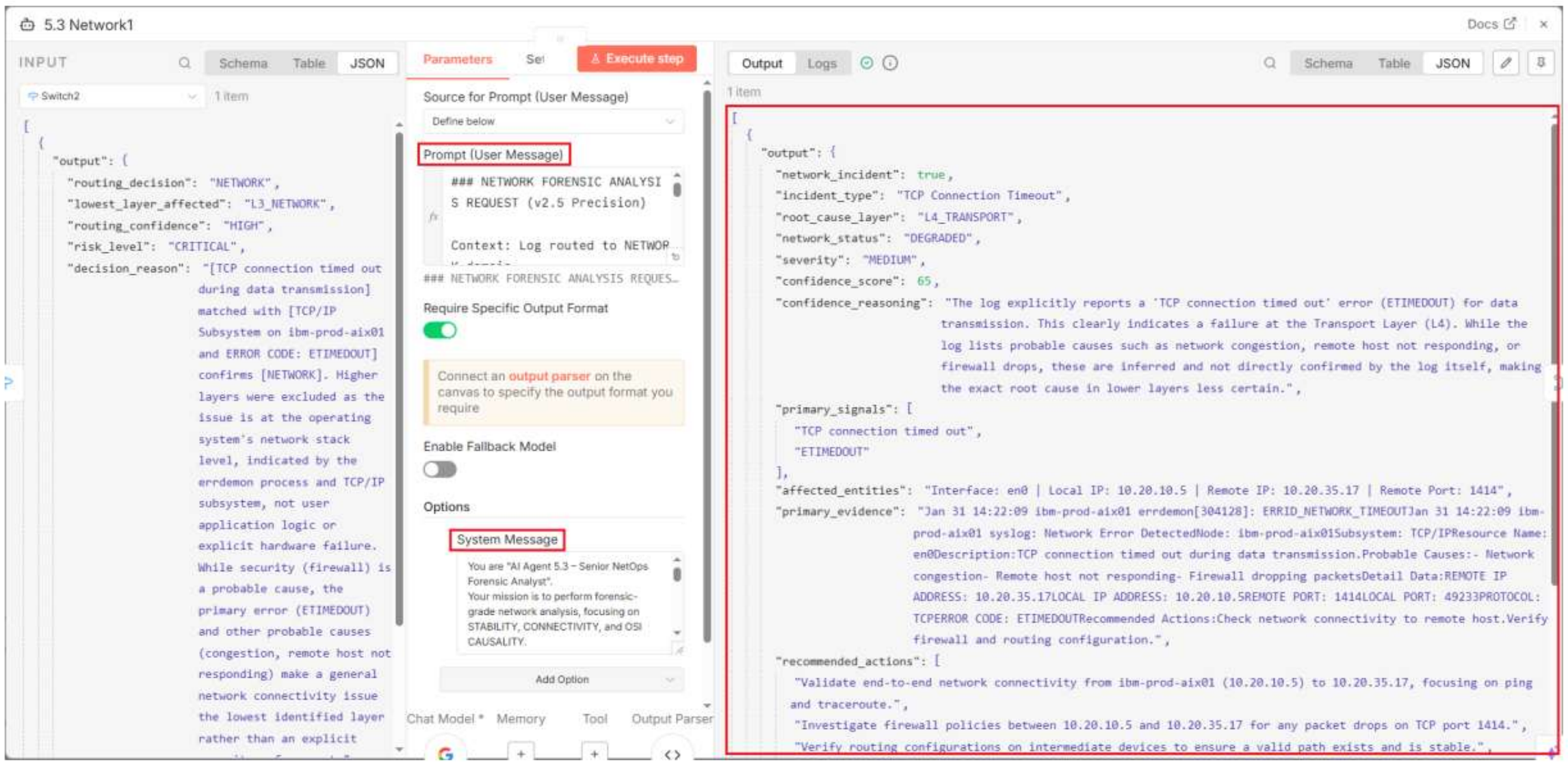

**Figure 11.** Network Evidence Extraction Workflow

*Application Runtime Analysis:* focuses on the Application Runtime Layer (see Figure 12). It targets to distinguish between Internal Code Failures (i.e., bugs, exceptions) and External Dependency Failures (i.e., database timeouts, upstream API errors). This distinction is critical for routing tickets to the correct team (Dev vs. SRE).

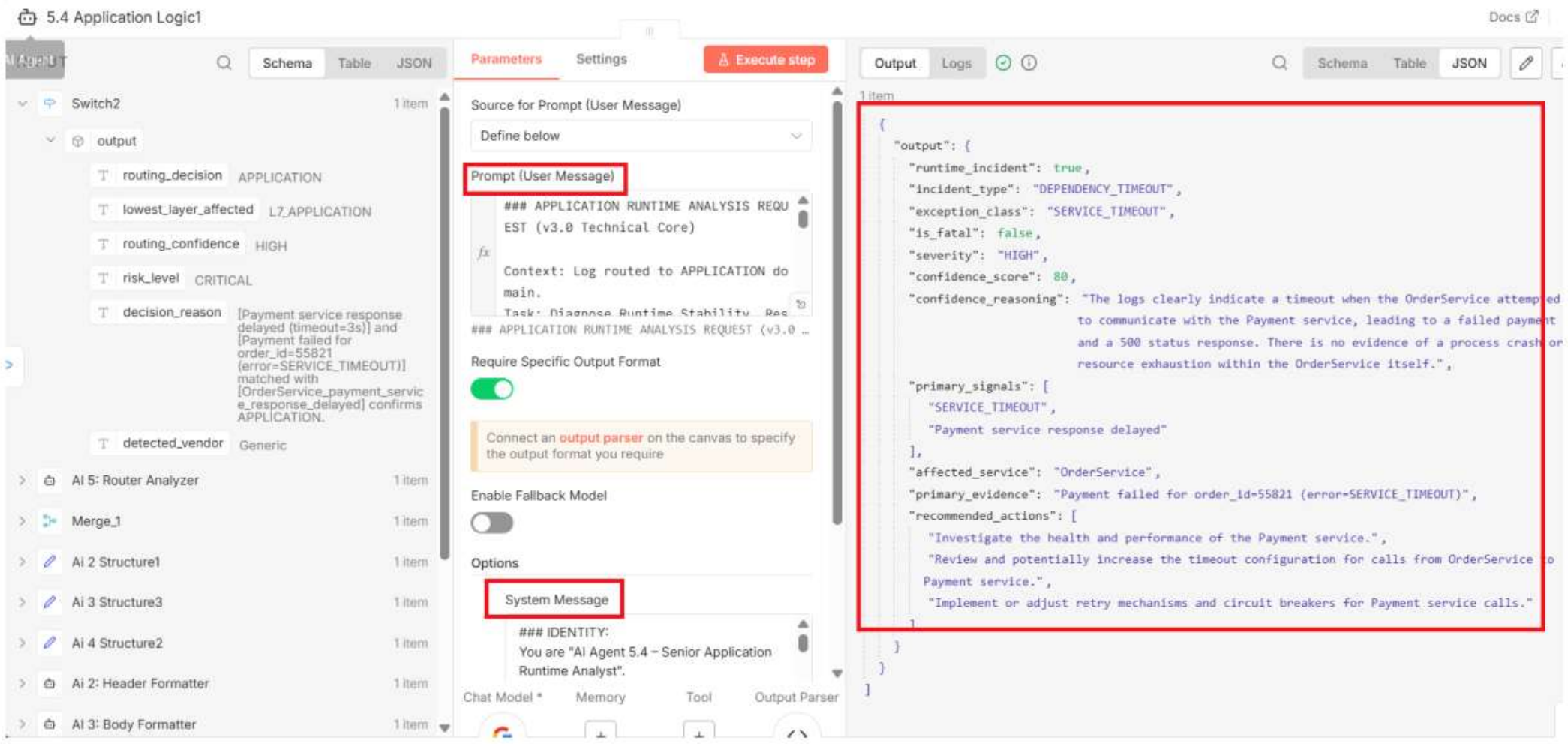

**Figure 12.** Application Runtime Configuration

## Hallucination Guard

This agent validates the analytical results against the raw log by using a binary PASS/FAIL logic (see Figure 13). It enforces three core integrity rules: 1) Verbatim check: the extracted evidence must match the raw log character-for-character. 2) Score governance: confidence scores must follow domain conditions, e.g., Blocked Attack ≤ 92%. 3) Domain isolation: agents are prohibited from diagnosing issues outside their assigned domain.

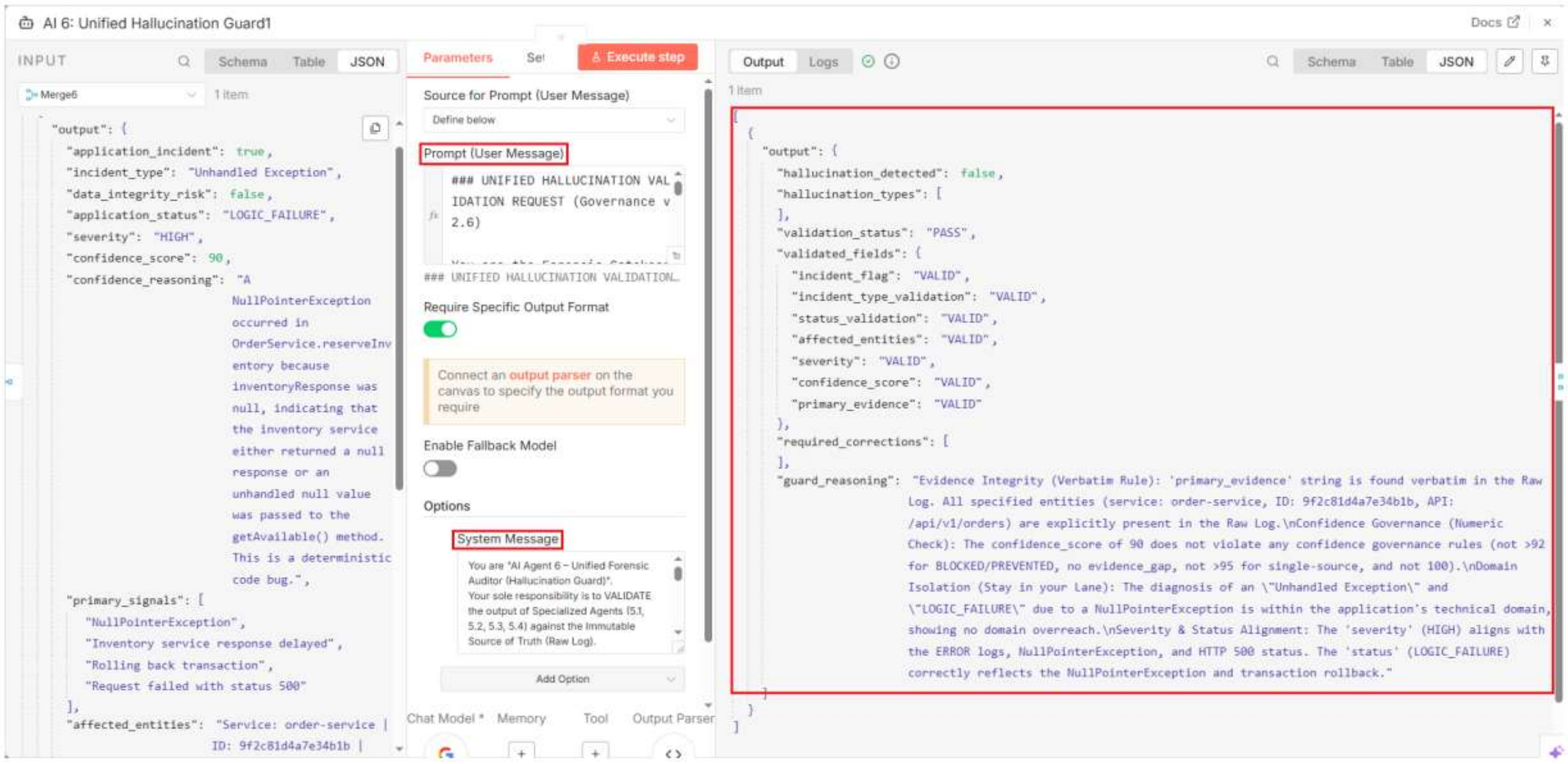

**Figure 13.** Hallucination Guard

## Aggregator

This agent checks the audit report from the analyzing agent (i.e., AI agent 6) and corrects problems (see Figure 14). If the evidence is missing or not reliable, the incident is rejected to avoid wrong conclusions. If the evidence is correct but the confidence score is too high (e.g., 99% for a blocked attack), this agent lowers the score to the allowed limit (such as 92%) instead of rejecting the incident. This method keeps valid security events while still following strict rules.

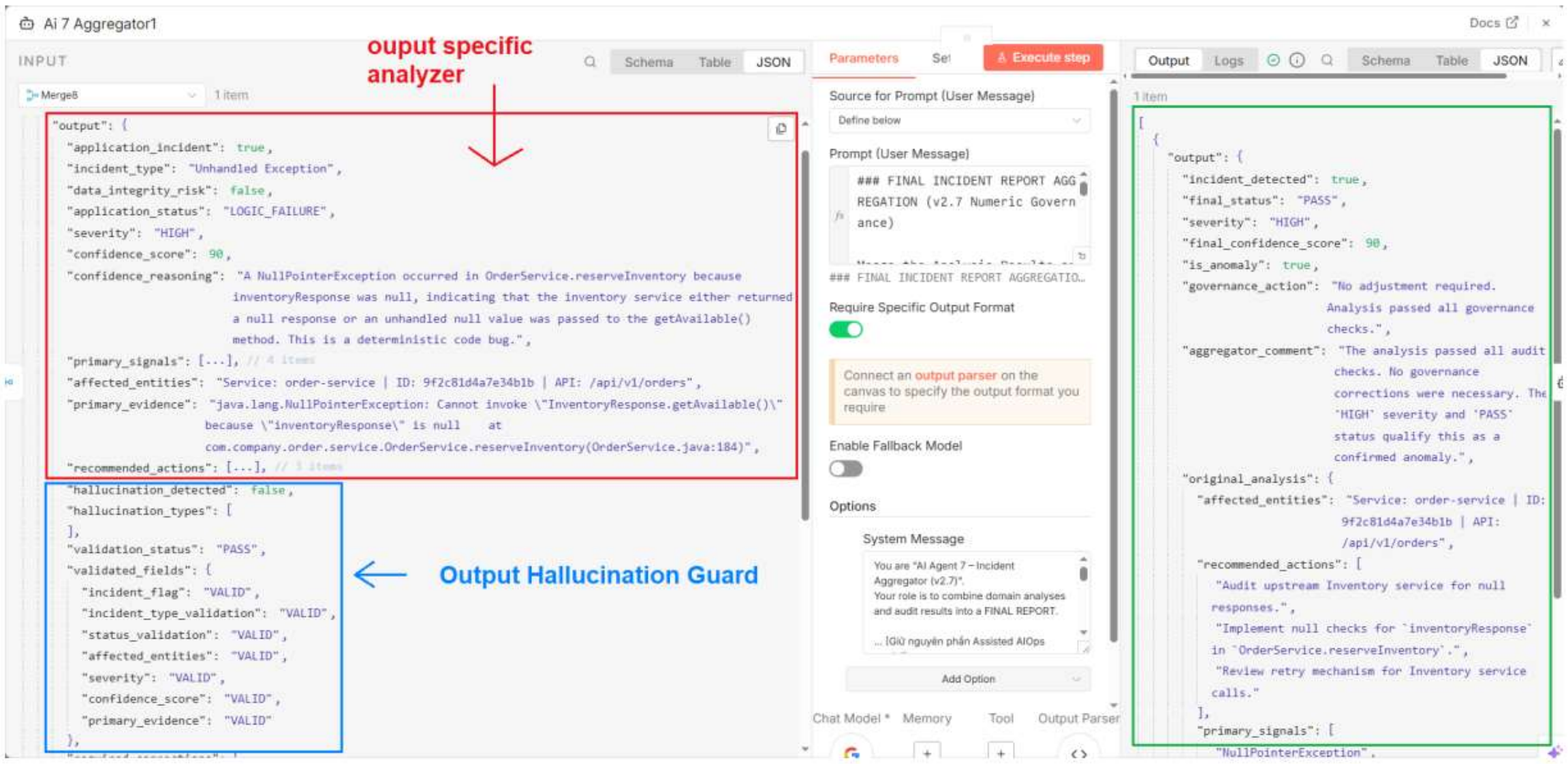

**Figure 14.** Aggregator Agent Configuration

# Results

**Test case 1**. We consider the real log error card by a Host Bus Adaptor (HBA) at a Dell Server. This case analyzes a hardware failure during system deployment. The lines 3-7 in Figure 15 shows an error alert.

```
1 2025-06-03 12:35:10,148Z INFO  Running <ImagingStepRAIDCheckPhoenix(Node
     10.30.3.34)>
2 2025-06-03 12:35:10,328Z INFO  Node is in Phoenix environment, generating
     hardware_config.json
3 2025-06-03 12:35:16,222Z ERROR Command '.local/bin/layout_finder.py local'
     returned error code 1
4 2025-06-03 12:35:16,223Z ERROR Failed to detect required storage controller HBA
5 Exception: This node is expected to have exactly 1 HBA with pci id 9005:028f,
     but Phoenix could not find any such device
6 2025-06-03 12:35:16,235Z ERROR hardware_config.json generation failed
7 2025-06-03 12:35:16,252Z DEBUG ImagingStepRAIDCheckPhoenix marked as FAILED
```

**Figure 15**. Dell Server Log for an HBA Failure

The result is outputted from the multi AI agent workflow as in Figure 16. The result demonstrates this is a critical error with a confidence rate 75% and confirms that the workflow detects an anomaly in the log with providing details. The recommendation for the fix is also presented in the result to the human admin.

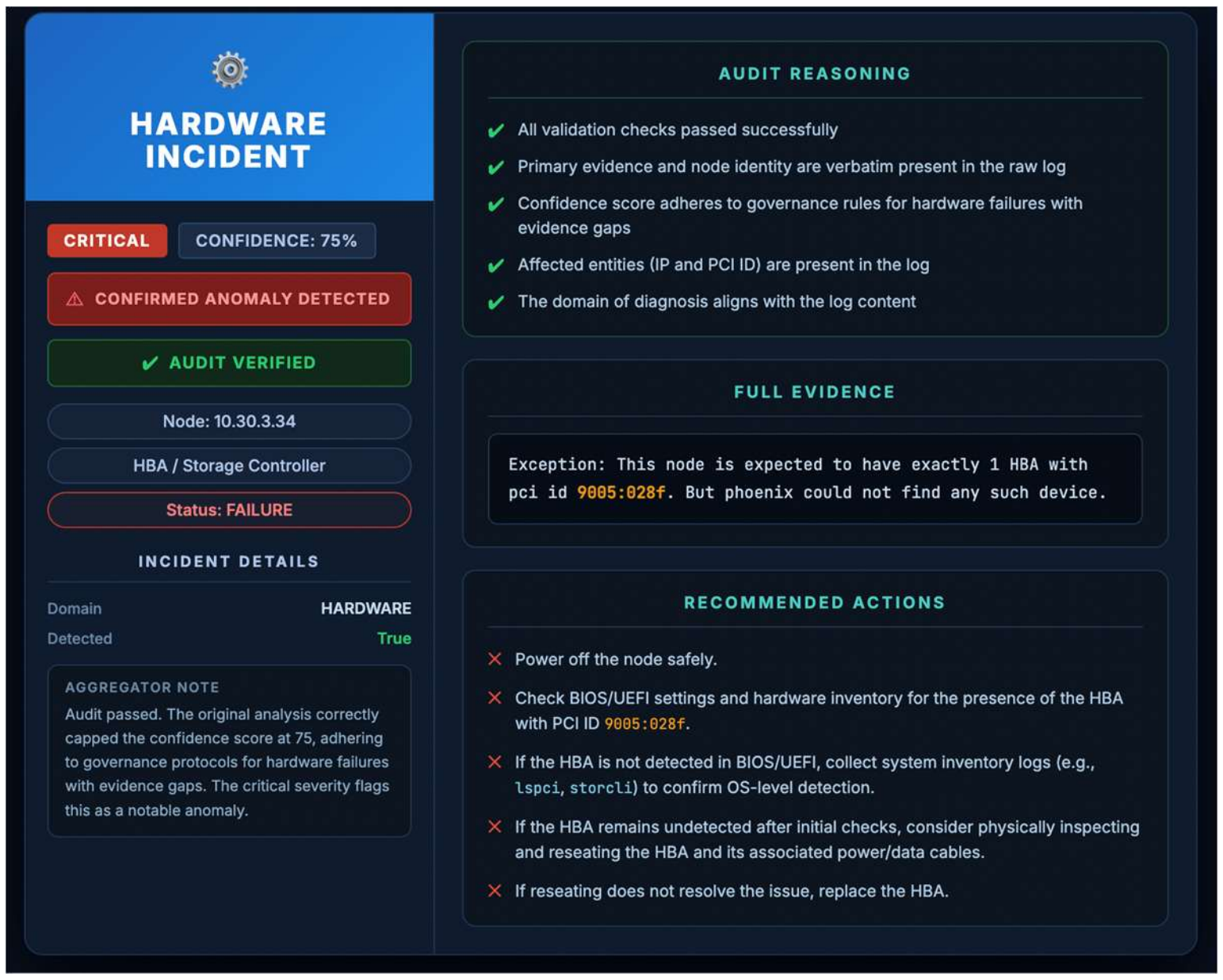

**Figure 16.** Results of HBA

**Test case 2.** We consider a log for a Cisco switch dead ports. The log content is presented in Figure 17. The error has been recorded on lines 7-11.

```
1  Cisco IOS Software, C2960X Software (C2960X-UNIVERSALK9-M), Version 15.2(7)E11
2  ...
3  Switch 1 booting as Master
4  Waiting for Port download...Complete
5  Initializing Port Extension Feature Support...
6  ...
7  Feb 27 09:59:08.876: %ILPOWER-3-CONTROLLER_PORT_ERR: Controller port error,
       Interface Gi1/0/1: Power Controller reports power supply VEE under voltage
       and lockout
8  Feb 27 09:59:08.876: %ILPOWER-3-CONTROLLER_PORT_ERR: Controller port error,
       Interface Gi1/0/2: Power Controller reports power supply VEE under voltage
       and lockout
9  Feb 27 09:59:08.880: %ILPOWER-3-CONTROLLER_PORT_ERR: Controller port error,
       Interface Gi1/0/3: Power Controller reports power supply VEE under voltage
       and lockout
10 ... [Repeated errors for multiple ports] ...
11 Feb 27 09:59:08.904: %ILPOWER-3-CONTROLLER_PORT_ERR: Controller port error,
       Interface Gi1/0/6: Power Controller reports power supply VEE under voltage
       and lockout
```

**Figure 17.** Cisco Switch Boot Log

This log was sent through the workflow and the output is presented in Figure 18. The error is reasoned to be critical and needs to be fixed at the confidence rate 75%. The error information is detailed in the result.

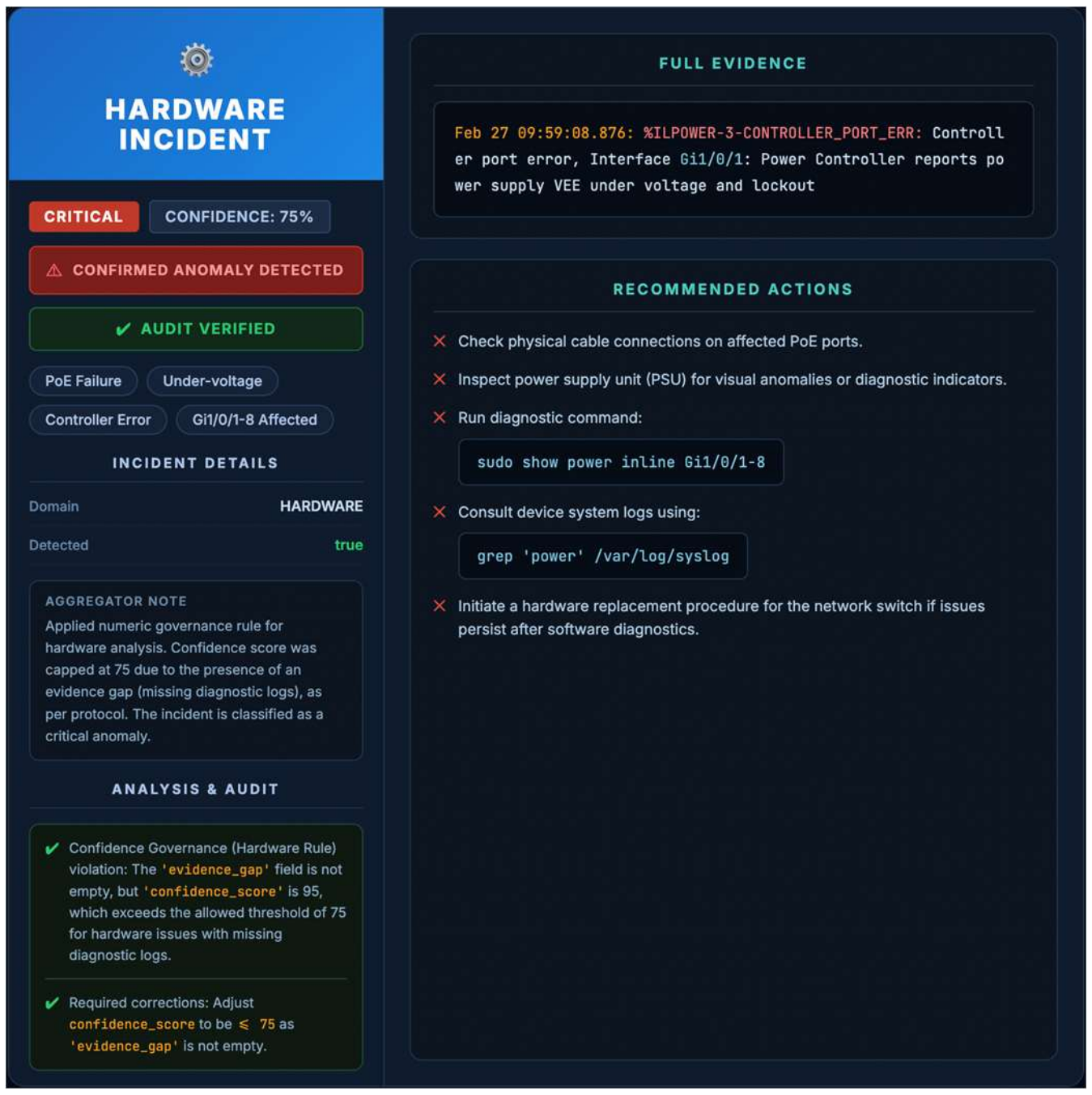

**Figure 18.** Results of Cisco switch dead ports

**Test case 3.** In this test case, we consider a log for an Apache HTTP error. The log is reported the matter as in Figure 19. This dataset contains nearly 4000 lines of logs (Seyyar et al., 2018). We focus on the most critical segments regarding the web operation.

```
192.168.4.163 - "GET /7Vi0ho7m.foobar HTTP/1.1" 404 507 "-"
192.168.4.163 - "GET /cdcZMNEA.cgi HTTP/1.1" 404 503 "-"
192.168.4.163 - "GET /index.php/component/users/?view=apyhv%3C%2F-%3E... HTTP
    /1.1" 500 2125 "-"
192.168.4.163 - "GET /index.php?format=feed&type=a%27b%22c%27d%22 HTTP/1.1" 500
    1983 "-"
192.168.4.163 - "GET /DVWA/dvwa/includes/dvwaPhpIds.inc.php HTTP/1.1" 200 224
    "-"
192.168.4.163 - "GET /index.php/component/mailto/?tmpl=... HTTP/1.1" 200 338 "-"
192.168.4.163 - "POST /DVWA/login.php HTTP/1.1" 302 385 "-"
192.168.4.163 - "POST /index.php/component/search/ HTTP/1.1" 500 1924 "-"
192.168.4.163 - "POST /index.php HTTP/1.1" 500 1924 "-"
192.168.4.163 - "GET /DVWA/setup.php HTTP/1.1" 200 1698 "-"
```

**Figure 19.** Extracted W3af Vulnerability Scanning Signatures

The log containing the error moves through the workflow and the results are outputted as in Figure 20. The web anomaly trace has been found in the log. It states that the severity level gets medium and the confidence rate gets 90% showing this is a high truth of a abnormal behavior of the monitored web system. For the details of the errors, the output clarifies on the result screen. The last AI agent in the workflow also recommends the action for the next step.

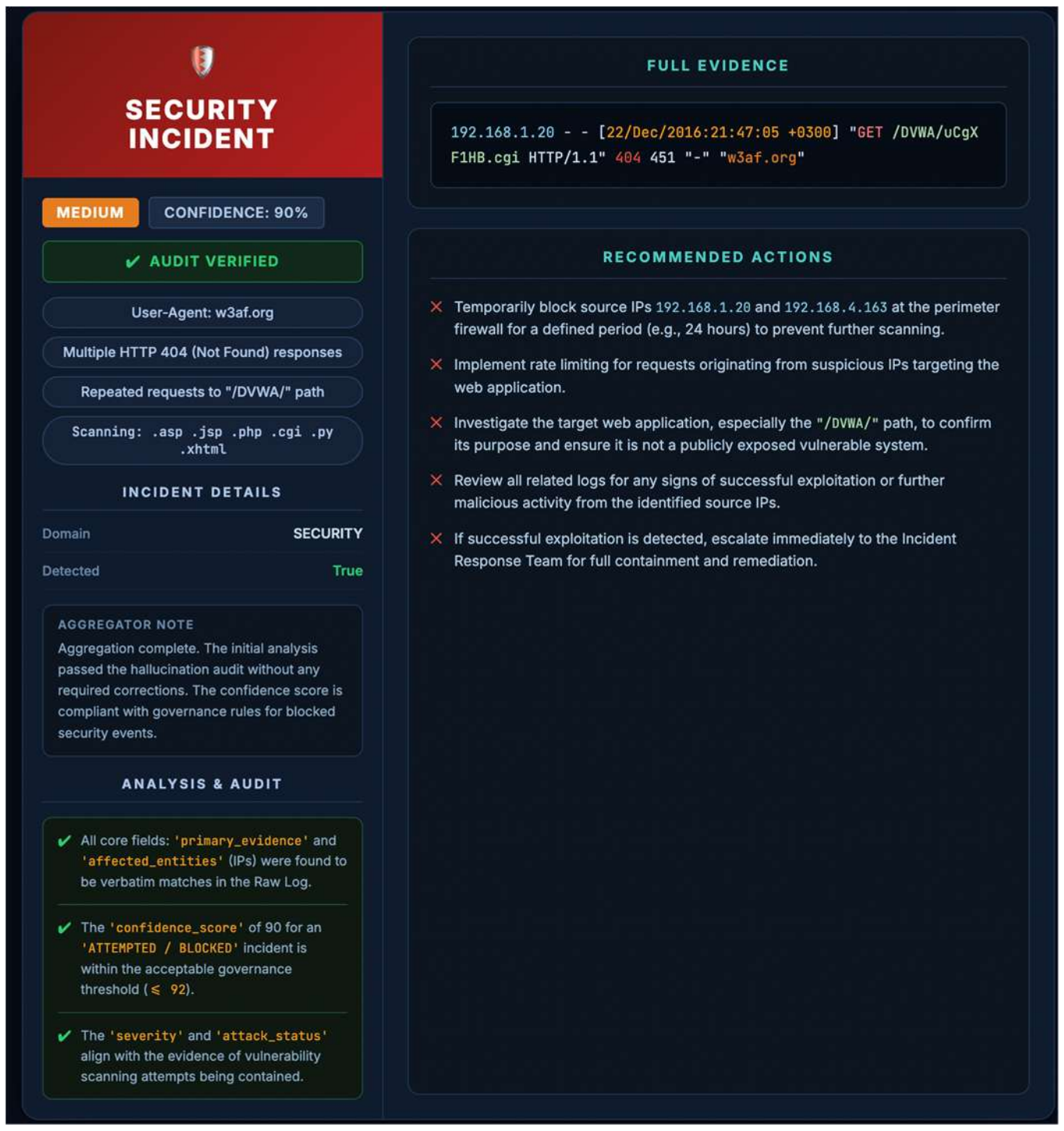

**Figure 20.s** Web Log Analysis Result

**Other test cases.** In addition to the 3 previous test cases, to evaluate the workflow performance and accuracy, we deployed 13 more test cases in 3 categories that is security (6 cases), system error (1 case), and application (6 cases). The more general information and results of these log analysis are described in Table 2. The logs are inputted to the agent 1.

**Table 2** List of test cases

| No | Category | Anomaly | Severity (%) | AI Confidence (%) |
|---|---|---|---|---|
| 1 | security_ops | true | Rather High | 80 |
| 2 | security_ops | true | High | 95 |
| 3 | security_ops | false | Mean | 50 |
| 4 | security_ops | true | High | 100 |
| 5 | security_ops | true | High | 100 |
| 6 | security_ops | true | High | 100 |
| 7 | system_infra | true | Low | 30 |
| 8 | app_logic | true | Rather high | 85 |
| 9 | app_logic | true | Rather high | 85 |
| 10 | app_logic | true | Rather high | 85 |
| 11 | app_logic | true | Rather high | 70 |
| 12 | app_logic | true | Rather high | 75 |
| 13 | app_logic | true | Rather high | 75 |

After inputting the logs in the workflow, as a consequence, each case's result shows a severity percentage showing the vulnerability level of the log and the AI confidence percentage for the accuracy of detecting the anomaly trace from the log. As in Table 2, the result of test case 3 for a security ops analysis is false. Being false on the anomaly column means the system event audited in the log is safe. In this case, the confidence for reasoning is 50%. This rate is not very high, just at the mean level. In the other test cases, such as with test cases 4-6, the severity is high and the AI confidence is very high. This means the logs in these test cases contain the anomalous trace very clearly, and the workflow can detect and point out correctly the forensic information, at the same time, its recommendation is reasonable. For the other test cases, the interpretation states that the performance of the workflow is very potential.

## Discussion and Conclusion

The paper shows that a multi agent AI framework operating from verbatim checks to confidence caps. The test cases show that the workflow operates effectively. However, the efficiency of the system depends heavily upon the quality of input logs to be analyzed. Moreover, the memory length used for each agent is still short. Therefore, in the future works, we will improve the automatic raw log process, and applying RAG techniques to handle complex, undocumented faults.